\documentclass[12pt,preprint]{aastex}

\newcommand{\um}{$\mu$m~}
\newcommand{\ums}{$\mu$m}

\shorttitle{ Spitzer Spectra of Faint Sources}
\shortauthors{Weedman et al.}

\begin{document}

\title{Spitzer IRS Spectra of Optically Faint Infrared Sources with Weak Spectral Features}

\author{D. W. Weedman\altaffilmark{1}, B. T. Soifer\altaffilmark{2,8}, Lei Hao\altaffilmark{1}, J. L. Higdon\altaffilmark{1}, S. J. U.
  Higdon\altaffilmark{1}, J. R. Houck\altaffilmark{1}, E. Le
  Floc'h\altaffilmark{5}, M. J. I. Brown\altaffilmark{3,4}, A.
  Dey\altaffilmark{4}, B. T. Jannuzi\altaffilmark{4}, M. Rieke\altaffilmark{5}, V. Desai\altaffilmark{8}, C. Bian\altaffilmark{8}, D. Thompson\altaffilmark{8}, L. Armus\altaffilmark{2}, H. Teplitz\altaffilmark{2}, P. Eisenhardt\altaffilmark{6}, S.P. Willner\altaffilmark{7}}

\altaffiltext{1}{Astronomy Department, Cornell University, Ithaca, NY 14853; dweedman@astro.cornell.edu}
\altaffiltext{2}{Spitzer Science Center, California Institute of Technology, 220-6, Pasadena, CA 91125}
\altaffiltext{3}{Department of Astrophysical Sciences, Princeton University, Peyton Hall, Princeton, NJ 08544-1001}
\altaffiltext{4}{National Optical Astronomy Observatory, Tucson, AZ 85726}
\altaffiltext{5}{Steward Observatory, University of Arizona, Tucson, AZ 85721}
\altaffiltext{6}{Jet Propulsion Laboratory, 4800 Oak Grove Dr., 169-327, Pasadena, CA 91109}
\altaffiltext{7}{Smithsonian Astrophysical Observatory, 60 Garden St., Cambridge, MA 02138}
\altaffiltext{8}{Division of Physics, Mathematics and Astronomy, 320-47, California Institute of Technology, Pasadena, CA 91125}

\begin{abstract}
  
Spectra have been obtained with the low-resolution modules of the Infrared Spectrograph (IRS) on the Spitzer Space Telescope ($Spitzer$) for 58 sources having f$_{\nu}$(24\,\um) $>$ 0.75\,mJy.  Sources were chosen from a survey of 8.2 deg$^{2}$ within the NOAO Deep Wide-Field Survey region in Bo\"{o}tes (NDWFS) using the Multiband Imaging Photometer (MIPS) on $Spitzer$.  Most sources are optically very faint ($I$ $>$ 24\,mag).  Redshifts have previously been determined for 34 sources, based primarily on the presence of a deep 9.7\,\um silicate absorption feature, with a median z of 2.2.  Spectra are presented for the remaining 24 sources for which we were previously unable to determine a confident redshift because the IRS spectra show no strong features. Optical photometry from the NDWFS and infrared photometry with MIPS and the Infrared Array Camera on $Spitzer$ (IRAC) are given, with $K$ photometry from the Keck I telescope for some objects.  The sources without strong spectral features have overall spectral energy distributions (SEDs) and distributions among optical and infrared fluxes which are similar to those for the sources with strong absorption features.  Nine of the 24 sources are found to have feasible redshift determinations based on fits of a weak silicate absorption feature. Results confirm that the "1 mJy" population of 24\,\um $Spitzer$ sources which are optically faint is dominated by dusty sources with spectroscopic indicators of an obscured AGN rather than a starburst.  There remain 14 of the 58 sources observed in Bo\"{o}tes for which no redshift could be estimated, and 5 of these sources are invisible at all optical wavelengths.

\end{abstract}



\section{Introduction}

Imaging surveys at infrared wavelengths with the Spitzer Space Telescope ($Spitzer$) have the potential to reveal populations of sources that were previously unknown.  In particular, surveys with the Multiband Imaging Photometer for
$Spitzer$ (MIPS) \citep{rie04} at 24\,\um should reveal sources which are luminous because of emission from dust, and which may contain sufficient dust to make them very faint or invisible optically.  The Infrared
Spectrograph on $Spitzer$ (IRS)\footnote{The IRS was a collaborative venture between Cornell
University and Ball Aerospace Corporation funded by NASA through the
Jet Propulsion Laboratory and the Ames Research Center.} \citep{hou04} is sufficiently sensitive to obtain low-resolution spectra of these MIPS sources at flux density levels of f$_{\nu}$(24\,\um) $<$ 1 mJy.  

As part of initial efforts to characterise this "1 mJy" population, we 
surveyed 8.2 deg$^{2}$ within the Bo\"{o}tes field of the NOAO Deep Wide-Field Survey
(NDWFS) \citep{jan99} with MIPS to produce a catalog of mid-infrared
sources.  The MIPS data were
obtained with an effective integration time at 24\,\um of $\sim$90\,s per sky
pixel, reaching a 5 $\sigma$ detection limit of $\sim$ 0.3\,mJy for unresolved sources.  
This field was chosen because the deep and well calibrated optical
imagery in $B_W$, $R$, and $I$ bands makes possible the identification of infrared
sources with very faint optical counterparts and allows confident selection of infrared sources lacking optical counterparts to very deep optical limits.  

We then selected for subsequent spectroscopic observations with $Spitzer$  those sources which are the faintest optically (typically $I$ $>$ 24) while also bright enough ($>$ 0.75\,mJy at 24\,\um) for spectroscopy with the IRS within integration times of $\sim$ 1 h. Other than the mid-infrared flux limit, the only selection criterion was optical faintness, because we were primarily interested in understanding sources which would not have been identified in previous optical studies.  Our first results reported the discovery of a significant population of optically obscured, high redshift sources (\citet{hou05}; hereinafter H05).  To date, we have observed 58 Bo\"{o}tes sources with the IRS.  Continuum was detected in all 58 objects observed, confirming that all mid-infrared sources are real even when they have no optical counterparts and also demonstrating that the MIPS-derived positional uncertainties
are less than $\sim$ 0.5'' rms. Redshifts for 17 sources determined by fitting templates of known local objects were reported in H05; an additional 17 sources with redshifts include 14 in Higdon et al. (in preparation), 2 in \citet{des06}, and one in \citet{dey05}.  The most important result is that these sources are generally at high redshift. Of the 34 sources with redshifts, the median redshift (z) is 2.2.  The spectra of these sources with redshifts are dominated by strong silicate absorption centered at rest frame 9.7 $\mu$m. Only two sources are best fit by strong PAH emission features not requiring silicate absorption.  There remain 24 spectra which we have observed with the IRS but for which redshifts have not yet been reported because there are no strong spectral features.  We discuss these sources in the present paper. 

All Bo\"{o}tes sources with 1.9 $<$ z $<$ 2.7 from IRS redshifts have measurable redshifts because of the deep silicate absorption feature centered at rest frame 9.7\,\um.  Because of the accessible wavelength range of the IRS, redshifts beyond about 2.8 could not be measured because this absorption feature would be longward of the IRS wavelength limit. This allows two possible interpretations of the 24 sources for which we have not been able to derive redshifts: either they are a category of sources at redshifts z $\la$ 2.8 having weak spectral features, or they are at z $\ga$ 2.8.  It is crucial to know which conclusion is more representative of the "no-z" sample.   If features are weak, that is important for interpreting the nature of the sources, and for understanding why they are optically faint.  If the "no-z" sources are at higher redshifts than for the measured sources, this would be evidence for an obscured population at even higher redshifts and luminosities than the sources already identified.  

Because this population of optically-faint infrared sources would not have been selected using criteria available before $Spitzer$, it is necessary to consider whether all of the targets do indeed represent a distant, extragalactic population, as we have assumed.  Our initial selection ruled out Solar System objects by verifying that the infrared source showed no proper motion between the two epochs of the MIPS observations; the selection at 24\,\um insures that any Solar System
objects would be in the main asteroid belt or closer and hence would have
measurable proper motion between $Spitzer$
observations.  All sources were also detected at the same positions with the $Spitzer$ Infrared Array Camera (IRAC) \citep{faz04} with observations  obtained at a different epoch.  Known stellar populations are ruled out because the extreme infrared to optical flux ratios imply very cool source temperatures for a black body.  In addition, we report new $K$ band observations with high spatial resolution which resolve at least 4 of the 24 sources. All indications, therefore, are that these sources are an extragalactic population, but locating them in the universe requires the determination of redshifts. 

In this paper, we give all available data for the 24 sources with "featureless" spectra and compare the overall spectral energy distributions for this sample of sources to those sources with previously determined redshifts.  We illustrate all of the spectra in order to discuss which ones might have weak features and suggest new redshifts for 9 sources based on possible but weak spectral absorption features. 

\section{Observations and Data Reduction}

Our initial selection for spectroscopic targets in the Bo\"{o}tes field was to 
use the MIPS and optical surveys to inspect all sources within the 8.2 deg$^{2}$ portion of the MIPS survey field overlapping the NDWFS which have f$_{\nu}$(24\,\um) $>$ 0.75 mJy and $I$ $\ge$ 24\,mag. There are 4273 MIPS 24\,\um sources brighter than 0.75\,mJy, of which 114 met this optical
magnitude criterion. (Magnitudes in the NDWFS Data Release 3 are slightly modified from those in the initial version from which we worked, so these statistics would not be exactly correct using the released version.)  There are 65 sources of the 114 having f$_{\nu}$ (24\,\um) $>$ 1.0\,mJy and $I$ $>$ 24\,mag.   To define a sample based on specific infrared and optical flux limits, we observed 35 of the 65 sources with f$_{\nu}$(24\,\um) $>$ 1.0\,mJy and $I$ $>$ 24\,mag.  We also included 23 sources with extreme values of infrared to optical flux ratio even if they did not meet these specific selection criteria. (In the remainder of this paper, we define the infrared to optical flux ratio as IR/opt $= \nu$f$_{\nu}$(24\,\um)$/ \nu$f$_{\nu}$($I$).) These remaining 23 targets are primarily sources with f$_{\nu}$(24\,\um) $<$ 1.0\,mJy and $I$ $>$ 24\,mag, but a few sources with optical magnitudes $I$ $<$ 24\,mag were also included. 

Spectroscopic observations were made with the IRS Short Low module in
order 1 only (SL1) and with the Long Low module in orders 1 and 2 (LL1 and
LL2), described in \citet{hou04}.  These give low resolution spectral
coverage from $\sim$8\,\um to $\sim$35\,\um.  Sources were normally placed on
the slits by offsetting from nearby 2MASS stars; in a few cases with no sufficiently nearby 2MASS stars, direct pointing without offsets was used successfully.

All images when the source was in one of the two nod positions on each
slit were coadded to obtain the source spectrum.  The background which was subtracted for LL1 and LL2 included coadded background images that added both nod positions with the source in the other slit together with the alternative nod position in the same slit, yielding a background observation with three times the integration time as for the source, in order to reduce noise in the background. 
The differenced source-minus-background image was used for the spectral extraction, giving independent extractions of the spectrum at the two positions on the slit for each LL order. The two independent spectra were compared to reject any highly outlying pixels in either
spectrum, and a final mean spectrum was produced.  For SL1, there was no separate background observation with the source in the SL2 slit, so background subtraction
was done between coadded images of the two nod positions in SL1.  Observed images were processed with version 11.0 of the SSC pipeline. Extraction of
source spectra was done with the SMART analysis package \citep{hig04}. All spectra discussed in this paper are shown in
Figures 1 and 2; displayed spectra have been boxcar-smoothed to the approximate resolution of the different IRS modules (0.2\,\um for SL1, 0.3\,\um for LL2, and 0.4\,\um for LL1).  It can be seen from comparison of the spectral flux densities with the MIPS fluxes (Figures 1 and 2) that the extracted spectra typically agree at 24\,\um to within 10 \% of the MIPS flux.   

Table 1 summarizes the MIPS and IRS observations and resulting characteristics for the 24 sources, giving the name (with full coordinates) and the MIPS 24\,\um flux
density. Sources are ordered by declination. Even when spectral features cannot be seen, the IRS spectra yield a slope of the continuum. To record this information, the power-law slopes of the continua are determined by using SMART to find the best linear fit to the continuum when displayed in log-log units.  On occasion with various IRS spectra, the SL continuum does not perfectly stitch to the LL continuum, and this is attributed primarily to slight mispointing, which can cause significant effects in the SL slit.  With spectra having poor signal-to-noise ratios (S/N), poor stitching might not be obvious, so the formal measures of the power law index listed in Table 1 use only LL spectra, with $\lambda$ $>$ 14\,\um. Fits are cut at 33\,\um because of the increased noise in the spectra that appears beyond this wavelength. The resulting power law index for 14\,\um $<$ $\lambda$ $<$ 33\,\um is given in Table 1.  We have also combined IRAC fluxes with the MIPS results to determine the slope of the power law which would connect the MIPS 24\,\um and IRAC 8\,\um fluxes as well as the power law connecting IRAC 8\,\um and 3.6\,\um fluxes.  These slopes are also given in Table 1. 

To quantify the quality of the spectra, we measure the S/N of the LL1 spectra.  This is done by determining within SMART the best-fit continuum to the LL1 spectrum of either linear, quadratic, or cubic fit and then measuring the RMS about this continuum. The S/N in Table 1 is the ratio S/N = (flux density at wavelength midpoint/RMS for full continuum fit).   Table 1 also includes any new redshift estimates that can be determined from all of the available data, as are discussed below.  

More details of near-infrared fluxes and optical magnitudes of the sources are in Table 2, including the IRAC flux measurements of these sources, observed in Bo\"{o}tes as described by \citet{eis04}, along with $K$ band magnitudes. 

The $K$ band data for 7 objects were obtained on UT
2005 June 19 and 20 with the Near Infrared Camera (NIRC) \citep{mat94} on the Keck I Telescope.  The NIRC instrument has a 38.4'' field of view and a pixel scale of 0.15''.  The seeing was
in the range 0.7'' to 0.9'' for all observations. Five objects were observed under
photometric conditions on June 20, each for a total exposure time of 20
minutes.  Objects number 4 and 24 were observed on June 19 under non-photometric
conditions for 16 and 21 minutes, respectively.  They were each
observed for an additional 200 seconds under photometric conditions on
June 20.

The individual dithered images for each target were dark-subtracted,
flat-fielded, masked, aligned, and stacked using the
IRAF\footnote{IRAF is distributed by the National Optical Astronomy
Observatory, which is operated by the Association of Universities for
Research in Astronomy, Inc., under cooperative agreement with the
National Science Foundation.} NIRCtools package. The data were placed on the Vega magnitude
scale using standard stars \citep{per98} observed at a range
of air masses to correct for extinction.  For objects 4 and 24, the
individual frames from June 19 were scaled to match the photometry
from June 20, and all frames were then combined using variance
weighting.

All seven objects were detected, with $18.6 < K < 20.9$ mag.  The
photometric uncertainties range from 0.1 to 0.3 mag and are dominated
by the source flux measurement, but include the zero point
uncertainties from the standard stars.  Sources 4, 9, 16 and 17 are clearly
resolved.

\section{Discussion}

\subsection{Characteristics of the Bo\"{o}tes Sample}

Understanding the full extragalactic population of mid-infrared sources is crucial to determining the evolution of dusty, obscured sources in the universe and to determining how those sources compare in number and evolution to optically visible sources.  Already there are indications from IRS results for the Bo\"{o}tes sources that the high redshift population of luminous, dusty galaxies which are optically obscured exceeds the population previously known from optical surveys.  H05 argue that the obscured sources with IRS redshifts determined from silicate absorption derive their mid-infrared luminosity primarily from an AGN, based on the high luminosities required, the similarity to infrared spectra of known local AGN, and the occasional presence of strong radio sources.  These IRS observations indicate that for redshifts z $\ga$ 2, the obscured quasars already discovered are a comparable population to unobscured quasars.  This conclusion derives from comparison with quasars discovered in the same Bo\"{o}tes survey at similar 24\,\um fluxes but which are optically bright. 

Using optically-determined spectroscopic redshifts, \citet{bro05} derive the redshift distribution and luminosity function of optically classifiable quasars discovered in the MIPS 24\,\um Bo\"{o}tes survey.  This result includes 87\% of the Bo\"{o}tes sources having f$_{\nu}$ (24\,\um) $>$ 1 mJy and $R$ $<$ 21.7 (for type 1 quasars, the $I$ mag limit would be similar).  This comprehensive sample has, therefore, a very similar 24\,\um selection criterion as our sample but much brighter optical magnitudes; the IR/opt for the Brown et al. sample is what is expected for unobscured, type 1 quasars.  Brown et al. conclude quantitatively that the shape of the rest-frame 8\,\um infrared luminosity function is the same as derived from optical surveys and that the redshift peak of quasar space density is at the same redshift.  In sum, selection of unobscured quasars based on 24\,\um fluxes gives the same result for comparable optical magnitude limits as a sample based on optical criteria alone. There are 35 quasars in the full Bo\"{o}tes field with 1.9 $<$ z $<$ 2.7 in their nearly-complete sample.

These numbers can be compared directly with the characteristics of our optically obscured sample within the same redshift interval by considering only those sources with 1.9 $<$ z $<$ 2.7 selected to the same infrared limit, having f$_{\nu}$(24\,\um) $>$ 1 mJy,  but with $I$ $>$ 24 mag.  There are 14 sources with IRS spectra observed in Bo\"{o}tes which meet these redshift and flux criteria.  There are a total of 65 objects in the full Bo\"{o}tes field satisfying both of these optical and infrared flux criteria, and 35 of these 65 were observed with the IRS.  Approximately, therefore, the 14 sources with redshifts 1.9 $<$ z $<$ 2.7, f$_{\nu}$(24\,\um) $>$ 1 mJy, and $I$ $>$ 24 mag derive from 35/65, or $\sim$ 54\%, of a complete sample defined by f$_{\nu}$(24\,\um) $>$ 1 mJy and $I$ $>$ 24.   Correcting for this factor of $\sim$ 2 incompleteness, there should be $\sim$ 27 optically-faint sources ($I$ $>$ 24 mag) having 1.9 $<$ z $<$ 2.7 in the full Bo\"{o}tes field.  This number is comparable to the 35 unobscured, type 1 quasars within this redshift range which are known in the field from \citet{bro05}.  This comparison gives an empirical indication that optically obscured sources already identified at these high redshifts are similar in number to the classical, optically discoverable type 1 AGN.  This is only a lower limit to the actual number of obscured sources, because the 24 objects in the Bo\"{o}tes sample without redshifts are not included, and partially obscured sources having f$_{\nu}$(24\,\um) $>$ 1 mJy in the intermediate magnitude range 22 $\la$ $I$ $\la$ 24 have not been counted in this tally of obscured AGN.  Our result that the number density of obscured quasars exceeds that of unobscured quasars at high redshifts is also consistent with estimates of obscured AGN derived using $Spitzer$ infrared colors compared with optical or radio classifications of AGN \citep{ah06, ms05}.  

To refine our measures of the obscured population, it is necessary to understand whether the 24 sources without IRS redshifts are also high redshift, obscured sources.  There are two alternative explanations for why strong spectral features are not visible in these 24 sources.  The first is that the redshifts are so high that features are shifted out of the range of the IRS. The highest redshift for which the silicate absorption feature is measured in the initial results of H05 is 2.73. If sources have redshifts above 2.8, the silicate absorption feature would be centered at observed wavelength of 37\,\um, so even the beginning of this feature could not be confidently identified in IRS spectra of poor S/N. There is one observed example already of a source with known redshift higher than this limit which shows a featureless spectrum within the IRS wavelengths.  This is the gravitationally lensed quasar APM 08279+5255, having z = 3.91, which shows a featureless power law over the full IRS range, corresponding to rest wavelengths of 
1.7\,\um to 7\,\um \citep{soi04}, with a power law slope of 1.5.  By comparison to this source, we would not expect to see features in sources with z $\ga$ 2.8.

The alternative explanation is that sources have redshifts which place the 9.7\,\um silicate absorption feature within our spectral range, but that this feature is too weak to be detected.  The S/N required to see such features depends on the strength of the feature, so we can never rule out the presence of sufficiently weak features in our spectra.  However, we had already determined that features at the level of the weakest silicate absorption feature in any of the AGN templates used for the redshift determinations in H05 are not present.  Templates used were Arp 220, IRAS F00183-7111, and Markarian 231; the weakest 9.7\,\um absorption feature within these templates is Markarian 231. For Markarian 231, the absorption depth is 50\% (the IRS spectrum is in \citet{wee05}); i.e., the absorption has removed 50\% of the unabsorbed continuum which would otherwise be present at the wavelength of maximum absorption. The absorption depth, defined as ABS = [1 - f(9.7)/f(CONT)], is estimated by comparing the observed flux density at 9.7\,\um, f(9.7), with the continuum flux density, f(CONT), predicted for an unabsorbed continuum as determined from a linear fit to the continuum level on  
either side of the silicate feature. The fraction of the continuum which is absorbed relates to the optical depth as ABS = 1 $-$ e$^{-\tau}$, for $\tau$ the maximum optical depth in the 9.7\,\um silicate absorption feature.  The six sources with the weakest features in H05 are fit with the Markarian 231 template which has ABS = 0.5; the remaining 11 sources have ABS $>$ 0.5.  

The detectability of silicate absorption depends on the value of ABS.  Of the 58 sources in Bo\"{o}tes observed in our faint-source sample, the 24 sources discussed in the present paper, or 41\%, are those with ABS $<$ 0.5; they did not have sufficient absorption to be fit with a template which would yield a confident redshift.  The sample of Bo\"{o}tes sources was chosen because of the extreme IR/opt, and we do not have a control sample of objects with known redshifts selected in the same way to determine what fraction of such sources would be expected to have deep silicate absorption.  The best set of such comparison sources currently available is a large sample of ultra-luminous infrared galaxies (ULIRGS) observed by the IRS \citep{spo06}. We have examined IRS low resolution spectra of 80 ULIRGS in this sample to estimate the depth of the 9.7\,\um silicate absorption feature (or height of any broad emission features, if present). Of the 80 sources, 60 (75\%) show the silicate absorption feature with absorption depth ABS $>$ 0.5 (Hao et al., in preparation). This result emphasizes the prevalance of deep absorption among ULIRGS and implies that absorption should be characteristic of dusty sources.  If the Bo\"{o}tes sources are all at redshifts such that the 9.7\,\um absorption feature is accessible to the IRS, we expect them to show absorption as frequently as do ULIRGS, given the Bo\"{o}tes selection designed to favor obscured sources.  That only 60\% of the Bo\"{o}tes sources show ABS $>$ 0.5 compared to 75\% of ULIRGS is not a statistically significant difference given the sizes of the two samples, so this comparison is consistent with (although does not prove) the interpretation that all of the Bootes sources are sources similar to absorbed ULIRGS.

Because Markarian 231 is being used as a comparison template for these sources, it is useful to ask how it would appear if at z = 2.  The redshift is 0.0422, and the observed f$_{\nu}$(8.4\,\um) = 1080 mJy \citep{rie76}, so the observed f$_{\nu}$(24\,\um) if at z = 2 would be 0.4 mJy.  The IR/optical is more uncertain because a total magnitude is not available at rest-frame 0.27\,\um for comparison to the observed $I$ band at z = 2.  The observed flux density at rest-frame 0.34\,\um is 4 mJy \citep{sur00}; the rest-frame flux density at 0.27\,\um for the nucleus alone, without correction for a total galaxy magnitude, is 0.5 mJy \citep{smi95}. These results bracket the true total magnitude at 0.27\,\um so yield that Markarian 231 would have 9 $<$ IR/opt $<$ 72 if at z = 2.  For scaling, a source of $I$ = 24 and f$_{\nu}$(24\,\um) = 1 mJy has IR/opt of 60.  The
sources in table 1 have IR/opt $\ge$ 40, and 75\% of these sources have
IR/opt greater than the maximum value that Markarian 231
could have.  Overall, therefore, Markarian 231 if at z = 2 would be too faint for our flux selection criterion for Bo\"{o}tes but might meet the IR/optical criterion.  It is sufficiently similar, however, to indicate that it is a valid comparison source.

\subsection{Comparison of Bo\"{o}tes Sources with and without Strong Spectral Features}

To consider whether the Bo\"{o}tes sources with weak spectral features are systematically different from the heavily absorbed sources with previously measured redshifts,  we compare the samples and look more closely at detailed spectral characteristics including overall SEDs.

The first test is to consider whether there are any obvious systematic differences between the Bo\"{o}tes sources with redshifts determined from deep absorption (ABS $>$ 0.5) compared to those sources in the present paper without strong spectral features.  Figure 3 shows the distributions of IR/opt and f$_{\nu}$(24\,\um) for the "no-redshift" sample in the present paper compared to the objects in H05 with redshifts.  The median f$_{\nu}$(24\,\um) is 1.1 mJy for those with redshifts and 1.3 mJy for those without. These flux limits are similar, implying that the ability to determine a redshift does not depend on flux. (In fact, observations were designed with the goal of having similar S/N at all fluxes, so that shorter integrations were used on brighter sources.) Counting the limits, sources with redshifts have median IR/opt of 145 and those without have 115.  This slight difference in the distributions for IR/opt implies that the sources without redshift may be somewhat less obscured optically compared to those with deep silicate features that allowed redshift determinations. However, the difference would correspond to an optical magnitude difference of only 0.25 mag, which is not a large difference compared to magnitude uncertainties.  In any case, the no-redshift sources are much more optically obscured than type 1 AGN, which is the category of AGN that does not typically show silicate absorption; the optically bright, type 1 quasars in the 1 mJy sample in \citet{bro05} have 1 $\la$ IR/opt $\la$ 10. The no-redshift sample remains puzzling, therefore.  How can sources which are seemingly very obscured optically have no strong absorption feature in the infrared?  

Another comparison of the samples with and without redshifts is through the near-infrared colors. Such colors have been used to preselect targets based on models of how colors derive for starbursts and AGN at various redshifts \citep{yan05} and in empirical comparisons of AGN and starbursts as classified by other criteria \citep{bra06}. Figure 4 shows the distributions of colors among MIPS 24\,\um and IRAC 8\,\um and 3.6\,\um fluxes.  There is little difference between the Bo\"{o}tes samples with and without redshifts.  The median log[f$_{\nu}$ (24\,\um)/f$_{\nu}$ (8\,\um)] is 1.0 for sources with redshifts, and 0.85 for sources without, indicating similar mid-infrared continuum slopes for both samples, which implies similar distributions of dust temperatures among the samples.

While this result does not distinguish the two samples, the slopes for both samples are steep enough to expect that many sources should be starbursts based on the spectral slopes.  \citet{yan05} considered a value of log[f$_{\nu}$(24\,\um)/f$_{\nu}$(8\,\um)]= 1.0 as dividing sources with starburst spectra ($>$ 1.0) from those with AGN spectra ($<$ 1.0).  By this criterion, about half of the objects in Figure 4 should be starbursts.  As previously noted, however, only 1 of the 24 Bo\"{o}tes sources in Figures 1 and 2 possibly shows PAH features, and it is these features which are associated with starbursts \citep{gen98}.

\citet{bra06} have recently demonstrated that AGN identified because of X-ray luminosity have a different distribution of mid-infrared colors compared to other mid-infrared sources.  They compare the IRAC and MIPS fluxes for all sources in the Bo\"{o}tes survey with the $\sim$ 5\% of sources detected by a Chandra X-Ray Observatory survey within the Bo\"{o}tes field.  Brand et al. show that the X-ray sources are distributed around a value of log[f$_{\nu}$(24\,\um)/f$_{\nu}$(8\,\um)] = 0.5, with a 1 $\sigma$ of 0.16 for the width of the distribution, whereas the remaining sources are distributed about a value of 1.0.  With this criterion, only half of the Bo\"{o}tes sources in Figure 4 fall within the 3 $\sigma$ distribution for an X-ray luminous AGN.  This again raises the question of why the Bo\"{o}tes sources with spectral slopes steeper than those of AGN rarely show evidence in the IRS spectra of the PAH features from starbursts.  

One explanation might be that these Bo\"{o}tes sources with steep slopes but no PAH features are AGN which have low X-ray luminosity and systematically cooler dust than the AGN associated with detectable X-ray sources.  An alternative explanation is that the sources with steep slopes but no PAH emission are not powered by AGN, but are powered by a starburst even though there is no PAH emission. In this circumstance, the starburst would have to be so embedded in dust or of sufficiently low metallicity that the PAH emission is absorbed or not produced, as suggested by \citet{hum05} to be characteristic of primordial starbursts, and as observed for the Blue Compact Galaxy SBS 0335-052 \citep{thu99, hou04b}.  Regardless of the explanation, the empirical conclusion is that the f$_{\nu}$(24\,\um)/f$_{\nu}$(8\,\um) does not give a consistent classification compared to mid-infrared spectral features for determining whether sources have their mid-infrared luminosity derived from an AGN or from a starburst.

Using the photometry from the NDWFS together with the IRAC and MIPS fluxes, the overall SEDs of the sources can be examined. Most of the sources in the Bo\"{o}tes sample have SEDs that are monotonically decreasing from the MIPS 24\,\um flux density through all four IRAC bands and the optical bands.  For such sources, these photometric continuum fluxes can be fit by a single power law or by a power law that becomes steeper at shorter wavelengths.  For the entire Bo\"{o}tes sample of 58 sources, 41 sources (71\%) have power-law spectra and so can be interpreted as having all infrared fluxes dominated by a dust continuum.  The remaining 17 sources show inflections among the IRAC points, indicating the possible presence of a stellar component.  For the 34 sources with redshifts, 25 are power laws. For the 24 no-redshift sources in the present paper, 16 are power laws. In this respect, the two samples are also similar.

These comparisons of IR/opt, mid-infrared colors, and SEDs do not indicate any differences in characteristics between the Bo\"{o}tes sources with strong absorption features and those with weak features.  Consequently, these comparisons do not indicate a conclusion regarding whether the no-z sources are at z $\la$ 2.8 or z $\ga$ 2.8.

Even though sources do not have sufficiently strong silicate absorption to be detected in our spectra, sources with weak silicate absorption could have sufficient optical absorption to produce large values of IR/opt.  As previously emphasized, the sources in the Bo\"{o}tes samples were selected to be optically faint, $I$ $\ge$ 24\,mag, and to have extreme values of IR/opt.  At redshift of 2, the rest-frame ultraviolet at 0.27\,\um observed in the $I$ band is heavily extincted compared to the extinction required to produce the 9.7\,\um absorption feature.  Using the grain model of \citet{li01}, the ratio of extinctions is A(0.27\,\um)/A(9.7\,\um) = 24, for A the extinction in magnitudes. A silicate absorption feature with ABS = 0.10, or optical depth of 0.1 in the 9.7\,\um feature, has A(9.7\,\um) = 0.11\, mag, so A(0.27\,\um) = 2.7\, mag.  This means that an unobscured, type 1 quasar of $I$ = 22\, mag would be extincted to $I$ $>$ 24\, mag if the ultraviolet continuum were absorbed by the same dust which produces a 9.7\,\um absorption feature of ABS = 0.10.  

A source with ABS = 0.10 would not have a sufficiently strong absorption feature to have been measured within the available Bo\"{o}tes spectra.  That such a source could nevertheless have significant optical extinction is consistent with explaining all of the Bo\"{o}tes sources as being obscured sources with large values of IR/opt arising from dust extinction, even though the 9.7\,\um feature was not detected in all of the IRS spectra. This interpretation would also explain why the IR/opt does not correlate with the detection of 9.7\,\um absorption; a source with negligible ABS can still have sufficient ultraviolet extinction to fall within a sample having $I$ $>$ 24\,mag.  This conclusion indicates that properties of the sources with weak silicate features are consistent with obscured AGN, but it is not proof of that classification.  It remains crucial to detect any weak absorption features within the spectra, which is discussed in the next section.

\subsection{Possible New Spectroscopic Redshifts}

To examine the spectra more carefully in a search for the possible presence of weak spectral features, we fit the power law that would connect the IRAC (8\,\um) and MIPS (24\,\um) photometry, and display this continuum in Figures 1 and 2.  By seeking departures from this power law, we can attempt to determine whether or not real spectral features are present. As previously emphasized, none of these sources are characterized by the readily visible, strong absorption features that allowed redshift measurements for other sources with similar S/N in H05. 
 
Sources are divided into Figure 1 and Figure 2 by placing in Figure 1 those which appear to have weak features that might be real.  The 12 spectra in Figure 1 have been fit with varying combinations of a power law and a silicate absorption profile \citep{li01}. Initially, the best fitting power law to the entire IRS spectrum is determined by adopting the least squares fit with the minimum residuals.  Then, the best fitting combination of a power law of varying slope with a silicate profile of varying depth is fit by stepping through all redshifts 1 $<$ z $<$ 3 in increments of 0.1. Results are compared for goodness of fit between the power-law fits and the power law plus absorption fits. For ten of the 12 sources in Figure 1, the fit containing absorption is quantitatively better.  An example is shown in Figure 5, for source 17.  The absorption fit shown has a redshift of 2.0 and ABS = 0.25. 

The sources in Figure 1 are now discussed individually, including comments on this absorption fit, and redshifts are listed in Table 1 when the fit using a weak silicate feature is statistically better than a power-law fit.

1. The spectral turnover starting at $\sim$25\,\um could be weak silicate absorption following the spectral peak at rest frame 8\,\um that is characteristic of absorbed sources; as seen in Markarian 231 and other absorbed ULIRGS, this spectral peak is not emission but arises because of absorption at longer and shorter wavelengths.  Fitting the spectrum with a power law and the silicate feature yields an optimum fit with ABS = 0.28 at z = 2.4.  

2. The spectral turnover starting at $\sim$23\,\um  could be absorption, which can be fit by a silicate feature of ABS = 
0.45 at z = 2.0, but the photometric z (discussed below) is 1.2.  The photometric z would be consistent with the broad peak centered at $\sim$20\,\um  being silicate emission at rest frame 10\,\um instead of the 8\,\um peak, but this would raise the puzzle of why an optically obscured source shows the silicate emission primarily found in unobscured, type 1 AGN. 

3. The possible absorption beyond $\sim$22\,\um can be fit by a silicate feature of ABS = 0.34 at z = 1.9. 

7. The possible absorption beyond $\sim$21\,\um can be fit by a silicate feature of ABS = 0.15 at z = 1.8, but this fit is not statistically better than a power-law fit without absorption, so a possible redshift is not listed in Table 1. 

9. The peak at $\sim$20\,\um could be the 8\,\um  peak, enhanced by noise, followed by very noisy absorption.  The $\sim$20\,\um peak could also be 7.7\,\um PAH emission at z $\sim$ 1.6, but we should then see the 11.2\,\um PAH feature at 29\,\um, which is not present.
A formal fit restricted to silicate absorption gives z $\sim$ 1.0. This spectrum is so noisy that no weight is given to this result, and the redshift of this object is considered as indeterminate. 

13. There is a reasonably convincing 8\,\um  peak at $\sim$24\,\um  followed by absorption, and this can be fit by a silicate feature of ABS = 0.30 at z = 2.2.

14. The peak at $\sim$20\,\um  could be the 8\,\um  peak, and the absorption can be fit by a silicate feature of ABS = 0.39 at z = 1.4.

15. The peak at $\sim$17\,\um  could be the 8\,\um  peak, and the absorption can be fit by a silicate feature of ABS = 0.37 at z = 1.1.

16. The fit with the standard silicate absorption profile fits the broader absorption beginning at 24\,\um, for a silicate depth of ABS = 0.29 and z = 2.0.  In this case, the narrower, deeper absorption at 21\,\um is attributed to a noise feature. 

17. This example of fitting the silicate feature is illustrated in Figure 5. The possible absorption beginning at $\sim$ 23\,\um  is fit by silicate absorption of ABS = 0.25 and z = 2.0. 

20. The peak at $\sim$21\,\um is reasonably convincing as being the 8\,\um peak followed by absorption, leading to a silicate fit of ABS = 0.27 and z = 1.6.  If this peak is instead the 7.7\,\um PAH emission, for z $\sim$ 1.7, the 11.2\,\um PAH emission would correspond to the weaker feature at 29\,\um , but the feature at 31\,\um would then have to be noise. 

22. The S/N is so poor for this source that the possible absorption beginning at the peak at $\sim$19\,\um cannot be fit by the power law plus silicate combination, so no redshift estimate is derived. 

Suggestions for new redshift assignments are listed in Table 1.  If we accept the suggested redshifts based on weak silicate absorption for the 9 sources in which this is identified (sources 1, 2, 3, 13, 14, 15, 16, 17, and 20), that brings to 43 the total number of spectroscopic redshift determinations for the 58 sources observed in Bootes.

There are a few objects in Figure 2 that might be questioned as to why we do not assign suggested features.  For source 6, little weight was given to the possible spectral turnover at $\sim$26\,\um  because the presence of absorption would depend on assuming that the sharp peak at $\sim$ 32\,\um represents continuum, but this peak is where noise features are often found. A similar objection applies to the apparent absorption at $\sim$ 30\,\um in source 12. For source 10, the S/N is so poor that the possible absorption at $\sim$26\,\um cannot be meaningfully fit. For source 11, the overall flatness of the spectrum resembles a PAH emission spectrum, but no consistent redshift arises from the apparent emission features, so some or all of these features must be noise features, and no redshift is assigned.

\subsection{Spectral Energy Distributions and Photometric Redshifts}

Even when features allowing a spectroscopic redshift determination are not present, it is sometimes possible to estimate a photometric redshift for faint infrared sources.  The estimate of a photometric redshift depends on identifying the redshifted peak in flux density at rest wavelength $\sim$1.6\,\um caused by the continuum maximum in giant or supergiant stars \citep{sim99}. For the 24 no-redshift sources in the present paper, 16 are power laws consistent with an AGN SED \citep{pol06}, so no photospheric peak is present, and no photometric redshift estimate is possible. There are 8 sources having IRAC inflections that deviate from a power law which can be fit with templates containing a stellar photospheric component. SEDs of these 8 sources are shown in Figure 6, with template fits which yield a photometric redshift. The template SEDs which are shown are intended only to be representative and to illustrate photometric estimates of redshift based on the 1.6u peak.  We did not attempt to determine models which optimize the SED fit or provide a quantitative separation of stellar luminosity and dust luminosity.

For most of the sources in Figure 8, the photospheric component is not conspicuous, presumably because a power-law component of dust continuum contributes to the flux at the wavelengths of the IRAC points and dilutes the photospheric component.  This is seen in Figure 6 for sources 1, 4, 6, 8, 11, and 16 for which the observed photometry at longer wavelengths exceeds the template.  Only for sources 2 and 12 do the observed fluxes at all wavelengths fit a full template including the photospheric component, so a photometric redshift estimate is more reliable for these two sources and is given in Table 1.   However, source 2 also has a redshift from a weak silicate feature (discussed above), which does not agree with the photometric redshift.  For the remaining 6 sources with evidence of a photospheric component, the redshift of the 1.6\,\um peak is much more uncertain, and photometric redshifts are not adopted.  Both of the photometric redshift assignments are at redshifts sufficiently low that the silicate absorption would be accessible to the IRS spectral window if present.

There is another interesting result regarding the 8 sources for which there is evidence for photospheric emission in the IRAC bands. All sources except source 8 have a ratio of f$_{\nu}$(24\,\um)/f$_{\nu}$(8\,\um) $>$ 9.5, and these 7 sources represent 7 of the 9 sources in the entire sample of 24 having f$_{\nu}$(24\,\um)/f$_{\nu}$(8\,\um) $>$ 9.5.  This limiting ratio is that used by \citet{yan05} with the intention of locating starburst sources.  While the SED fits are consistent with a substantial starburst component, we have already noted that only one of these sources (source 11) has an IRS spectrum that might conceivably show PAH emission.

\section{Summary and Conclusions}

We present IRS spectra and optical and infrared photometric data for 24 optically faint sources with f$_{\nu}$ (24\,\um) $>$ 0.8\,mJy selected from within the 8.2 deg$^{2}$ $Spitzer$ Bo\"{o}tes survey within the NOAO Deep Wide-Field Survey. The Bo\"{o}tes sample with IRS spectra comprises a set of 58 sources with high infrared to optical ratios ($\nu$f$_{\nu}$(24 \ums)$/ \nu$f$_{\nu}$($I$) = IR/opt $>$ 50) and typical f$_{\nu}$(24\,\um) = 1 mJy.  Redshifts have been previously determined for 34 of the 58 sources, with a median z = 2.2.  The 24 sources discussed in the present paper are the remaining sources which do not show sufficiently strong spectral features in the IRS low-resolution spectra for confident redshift determination by fitting with templates containing either strong silicate absorption or strong PAH emission.    

There are no significant systematic differences in overall SEDs or fluxes for the sample with redshifts (usually having strong silicate absorption at rest frame 9.7\,\um) and the sample without redshifts (having weak or no absorption or emission features).  16 of the 24 sources show power-law SEDs determined from photometry through infrared and optical wavelengths.  Sources could be at z $\la$ 2.8 with weak spectral features, or could show no features in the spectrum because z $\ga$ 2.8 and features are redshifted out of the IRS spectral range. 10 of the 24 sources have log [f$_{\nu}$(24 \ums)/f$_{\nu}$(8 \ums)] $>$ 1.0, a value expected for starbursts, but none of the sources show PAH emission features in the infrared spectra normally associated with a luminous starburst. 

Possible redshifts are suggested for 9 of the 24 sources based on fitting a profile of weak silicate absorption.  Photometric redshift estimates are given for 2 sources whose SEDs show evidence of a stellar component exceeding the dust continuum, although neither of these sources shows PAH emission, and the photometric redshift does not agree with the weak-silicate redshift for the one source with both measured. These suggested new redshifts provide evidence in favor of the explanation for the majority of the "no-redshift" sources that they are similar in nature and luminosity to the more heavily absorbed sources, but with weaker absorption features, and would bring the total number of redshifts determined for faint Bo\"{o}tes sources to 44 of the 58 sources observed, counting one photometric redshift.  Even if the suggested new redshifts are correct, there remain 14 sources from the Bo\"{o}tes sample of 58 with no redshift estimates.  With our present data, it is not possible to reach any conclusions regarding the nature of these remaining sources.  Five of these 14 sources without redshifts are not detected at any optical wavelength in the NDWFS.  Whether these represent a dusty population at redshifts higher than previously measured, or whether they represent a population of optically-obscured sources at z $<$ 2.8 with weak spectral features remains ambiguous. Because such sources are a significant fraction of the optically faint sources in the Spitzer "1 mJy" population of 24\,\um sources, continued efforts to determine their nature are important.

\acknowledgments
We thank D. Devost, G. Sloan and P. Hall for help in improving our IRS spectral analysis. We thank the staff of the Keck Observatory for their assistance with 
obtaining
the 2.2 micron observations. The W.M. Keck Observatory is operated as a 
scientific
partnership between the California Institute of Technology, the 
University of
California and NASA.  It was made possible by the generous financial 
support of
the W.M.Keck Foundation. We extend special thanks to those of Hawaiian 
ancestry
on whose sacred mountain we are priveleged to be guests. Without their 
generous
hospitality, none of the Keck observations presented herein would have been 
possible. We thank NOAO for supporting the NOAO Deep Wide-Field  
Survey; AD and BJ acknowledge support from NOAO, which is operated by  
the Association of Universities for Research in Astronomy (AURA),  
Inc., under a cooperative agreement with the National Science  
Foundation.  This work is based primarily on observations made with the
Spitzer Space Telescope, which is operated by the Jet Propulsion
Laboratory, California Institute of Technology under NASA contract
1407. Support for this work by the IRS GTO team at Cornell University was provided by NASA through Contract
Number 1257184 issued by JPL/Caltech.

\clearpage

\begin{deluxetable}{lccccccc} 
\tablecolumns{8}
\tabletypesize{\footnotesize}
\tablecaption{Observations and Properties of Sources}

\tablewidth{0pc}
\tablecaption{Observations and Properties of Sources}
\tablehead{
  \colhead{Source} & \colhead{Name\tablenotemark{a}} & \colhead{f$_{\nu}$(24\ums)\tablenotemark{b}} & \colhead{time\tablenotemark{c}} & \colhead{IR/opt\tablenotemark{d}}& \colhead{$\alpha$\tablenotemark{e}}&\colhead{z\tablenotemark{f}}&\colhead{S/N\tablenotemark{g}}\\
  \colhead{}&  \colhead{} &\colhead{(mJy)}& \colhead{(sec)}&\colhead{}&\colhead{}&\colhead{}&\colhead{}}
\startdata

1.  & SST24 J142936.71+323857.1 &  1.27  &1200,240 & 140 & 2.5, 2.2, 1.7&2.4,\nodata&3.6\\  
2.  & SST24 J143102.26+325152.3 &   1.34 &1200,240 & $>$180 & 0.9, 2.5, 1.4&2.0,[1.2]&5.1\\    
3.  & SST24 J143026.05+331516.4 &   1.83&960,240 & 60 & 2.4, 1.0, 0.9&1.9,\nodata&4.5\\
4.  & SST24 J143318.59+332127.0  &   0.91&1680,240 & $>$120 & 3.0, 2.4, 1.0&\nodata,\nodata&4.1\\
5.  & SST24 J142644.34+333052.0 &   1.12&1440,240 & 50 & 2.2, 1.1, 1.3&\nodata,\nodata&4.5\\
6.  & SST24 J143308.62+333401.7  &   2.40&720,240 & 270 & 2.7, 2.9, 1.8&\nodata,\nodata&6.8\\
7.  & SST24 J143053.14+334332.3 &   0.87 &1680,240 & $>$120  & 1.9, 1.6, $-$0.6&\nodata,\nodata&1.8\\
8.  & SST24 J142920.47+334400.7  &   0.92&1680,240 & $>$120 & 3.3, 1.5, 1.7&\nodata,\nodata&4.1\\
9.  & SST24 J143424.50+334543.3  &   0.88 &1680,240 & 50 & 2.4, 1.9, 1.7&\nodata,\nodata&2.8\\
10. & SST24 J142827.95+334550.3 &   0.82 &1920,240 & 100 & 2.0, 1.1, 0.9&\nodata,\nodata&2.9\\
11. & SST24 J143253.39+334844.3 &   1.40&1200,240 & $>$190 & 3.8, 2.1, 0.4&\nodata,\nodata&4.0\\
12. & SST24 J143004.77+340929.9 &   1.36&720,240 & 40 & $-$0.9, 3.2, 2.9&\nodata,[0.8]&4.1\\
13. & SST24 J143807.96+341612.5 &   2.19 &720,180 & $>$300 & 2.6, 2.3, 1.8&2.2,\nodata&4.7\\
14. & SST24 J142745.88+342209.0 &   1.52 &1680,240 & 110 &  3.0, 1.8, 1.8&1.4,\nodata&5.8\\
15. & SST24 J142842.96+342409.9 &   3.09  &480,120 & 80 & 2.1, 1.2, 1.0&1.1,\nodata&7.7\\
16. & SST24 J143429.56+343633.1  &   2.29 &720,240 & 110 & 3.1, 2.8, 2.0&2.0,\nodata&6.5\\
17. & SST24 J142940.85+344048.7 &   1.02 &1440,240 & 80 & 2.7, 2.2, 1.3&2.0,\nodata&5.0\\
18. & SST24 J142748.47+344851.3 &   2.10 &720,180 & 50 & 4.2, 1.4, 1.5&\nodata,\nodata&5.8\\
19. & SST24 J143213.40+350802.1 &   1.24  &1200,240 & $>$170 & 2.1, 1.2, 1.2&\nodata,\nodata&1.8\\
20. & SST24 J142611.38+351218.0 &   1.55 &1200,240 & 50 & 2.5, 1.4, 0.7&1.6,\nodata&3.1\\
21. & SST24 J142759.92+351243.5 &   1.45 &720,0& $>$200 & 2.4, 1.0, 1.1&\nodata,\nodata&4.8\\
22. & SST24 J143546.13+352447.2  &   0.99  &1440,240 & $>$130 & 2.1, 0.8, 1.0&\nodata,\nodata&2.8\\
23. & SST24 J142850.94+353146.6 &   0.98&1440,240 & $>$130 & 3.2, 1.6, 1.4&\nodata,\nodata&4.5\\
24. & SST24 J142939.18+353558.4 &   1.05  &1440,240 & $>$140 & 2.7, 0.9, $-$0.1&\nodata,\nodata&3.9\\



\enddata

\tablenotetext{a}{SST24 source name derives from discovery with the
  MIPS 24$\mu$m images; coordinates listed are J2000 24$\mu$m positions with typical 3\,$\sigma$ uncertainty of $\pm$ 1.2\arcsec. }

\tablenotetext{b}{Values of f$_{\nu}$(24 \ums) are for an unresolved point source, measured from MIPS images.}

\tablenotetext{c}{First number is total integration time for each order of the Long Low
  spectrum; second number is total integration time in Short Low order 1.}

\tablenotetext{d}{IR/opt $= \nu$f$_{\nu}$(24 \ums)$/ \nu$f$_{\nu}$($I$). For Vega magnitudes, zero magnitude in I band corresponds to 2450 Jy. 
  Limits are based on assumed $I$ $>$ 25. Values of $I$ are listed in Table 2.}

\tablenotetext{e}{$\alpha$ is the power law index for different portions of the continuum of form $f_\nu \propto \nu^{-\alpha}$; first number is the power law index that would connect the IRAC f$_{\nu}$(8 \ums) with the IRAC f$_{\nu}$(3.6 \ums); second number is the power law index that would connect the MIPS f$_{\nu}$(24 \ums) with the IRAC f$_{\nu}$(8 \ums); third number is the $\alpha$(14-33) for the continuum as determined from the best fit to the IRS spectra from 14\,\um to 33\,\um.}

\tablenotetext{f}{Redshifts are possible redshifts derived from fits of a weak silicate feature in the IRS spectra.  Redshifts in brackets are estimated from a photometric z based on inflection among the IRAC fluxes that can be fit with a template SED containing the 1.6\,\um peak in the spectrum of an evolved stellar component, shown in Figure 6.} 

\tablenotetext{g}{Signal to noise as measured in LL1 from the best-fit continuum longward of 20\,\um  by S/N = [f$_{\nu}$(27 \ums)/RMS], for RMS the 1$\sigma$ noise about the best-fit continuum.}		

\end{deluxetable}

\clearpage

\begin{deluxetable}{ccccccccc} 
\tablecolumns{9}
\tabletypesize{\footnotesize}
\tablecaption{Near-Infrared Flux Densities and Optical Magnitudes for Sources}

\tablewidth{0pc}
\tablecaption{Near-Infrared Flux Densities and Optical Magnitudes for Sources}
\tablehead{
  \colhead{SST24 Source Name}& \colhead{f$_{\nu}$(3.6\ums)\tablenotemark{a}}  &  \colhead{f$_{\nu}$(4.5\ums)\tablenotemark{a}} & \colhead{f$_{\nu}$(5.8\ums)\tablenotemark{a}}  &  \colhead{f$_{\nu}$(8\ums)\tablenotemark{a}} & \colhead{$B_W$}\tablenotemark{b}
  & \colhead{$R$}\tablenotemark{b}  & \colhead{$I$}\tablenotemark{b} & \colhead{$K$}\tablenotemark{c}\\
  \colhead{} &\colhead{($\mu$Jy)} &\colhead{($\mu$Jy)} &\colhead{($\mu$Jy)} &\colhead{($\mu$Jy)} & \colhead{(mag)} & \colhead{(mag)} & \colhead{(mag)}& \colhead{(mag)}}
\startdata

1.  J142936.71+323857.1  &15.8 &27.6 &33.4 &114 & 26.5    & 25.2    & 24.8  & \nodata \\  
2.  J143102.26+325152.3 &43.7 &41.2 &65 &90 &\nodata & \nodata & \nodata & \nodata\\    
3.  J143026.05+331516.4  &91 &162 &358 &622 & 24.7:    & 24.3:    & 23.4: & \nodata  \\
4.  J143318.59+332127.0  &6.0 &7.3 &33.1 &68.4 & \nodata & \nodata & \nodata &\nodata \\
5.  J142644.34+333052.0  &60.4 &84 &161 &353 & 26.4:    & 24.5:    & 23.8:  & 18.6 \\
6.  J143308.62+333401.7  &11.1 &14.3 &52.1 &96 & \nodata & 25.3    & 24.8 & 20.2  \\
7.  J143053.14+334332.3  &31.7 &48.4 &84.6 &142 & 26.4    & \nodata   & \nodata & \nodata \\
8.  J142920.47+334400.7  &12.2 &22.9 &8.3 &170 & \nodata & \nodata   & \nodata & 20.9  \\
9.  J143424.50+334543.3  &15.7 &21.1 &79 &107 & \nodata & \nodata & 24.0: & 19.8  \\
10. J142827.95+334550.3  &54.0 &83 &139 &257 & 25.4    & 25.5    & 24.9 & \nodata \\
11. J143253.39+334844.3  &6.5 &6.6 &21.5 &133 & \nodata & \nodata & \nodata & \nodata\\
12. J143004.77+340929.9  &85 &50.6 &38.7 &39.8 & 26.0   & 23.9    & 23.2 & \nodata\\
13. J143807.96+341612.5 &21.0 &28.1 &72 &168 & \nodata   & 25.5 & \nodata & \nodata \\
14. J142745.88+342209.0  &18.6 &39.9 &98 &212 & 24.8   & 24.2 & 24.3  & \nodata \\
15. J142842.96+342409.9  &155 &222 &413 &850 & 24.5   & 23.5   & 23.2  & \nodata \\
16. J143429.56+343633.1  &8.6 &8.8 &11.8 &106 & 26.4: & \nodata  & 23.9: & 20.7 \\
17. J142940.85+344048.7  &10.2 &16.5 &83 &89 & \nodata & \nodata  & 24.4 & 20.7 \\
18. J142748.47+344851.3  &16.1 &59.5 &157 &458 & 23.4    & 23.3     & 23.1  & \nodata\\
19. J143213.40+350802.1  &58.7 &95 &188 &323 & 26.4 & \nodata  & \nodata & \nodata  \\
20. J142611.38+351218.0  &48.0 &76 &172 &350 & 25.1   & 23.8     & 23.3 & \nodata  \\
21. J142759.92+351243.5  &67 &111 &239 &463 & 25.1    & 24.4     & \nodata & \nodata \\
22. J143546.13+352447.2 &71 &111 &188 &395 & 25.4    & 26.0     & \nodata & \nodata  \\
23. J142850.94+353146.6  &13.9 &29.4 &87 &179 & \nodata & \nodata  & \nodata & \nodata \\
24. J142939.18+353558.4  &43 &85 &168 &371 & \nodata & \nodata  & \nodata &20.1 \\

\enddata

\tablenotetext{a}{Fluxes from IRAC measures, assuming a point source.}
\tablenotetext{b}{Sources with an optical counterpart
  appear in NDWFS catalogs with prefix NDWFS and the optical source position; NDWFS Data Release 3 (DR3) is available at http://www.archive.noao.edu/ndwfs/. Magnitudes are Vega magnitudes for an unresolved source, measured from images in NDWFS DR3 by smoothing all images to a PSF of FWHM 1.35'' and applying  photometry at the 24$\mu$m source position with an aperture of 4'' diameter, except for objects with magnitudes followed by colon which use 3'' apertures because of proximity to a brighter source. The aperture photometry has been corrected to the total magnitude for an unresolved source by a correction of 0.1 mag brighter for the 4'' aperture and 0.3 mag brighter for the 3'' aperture.}
\tablenotetext{c}{$K$ magnitudes are total magnitudes. }

\end{deluxetable}

\clearpage
%
%
\begin{figure}
\figurenum{1}
\includegraphics[scale=0.7]{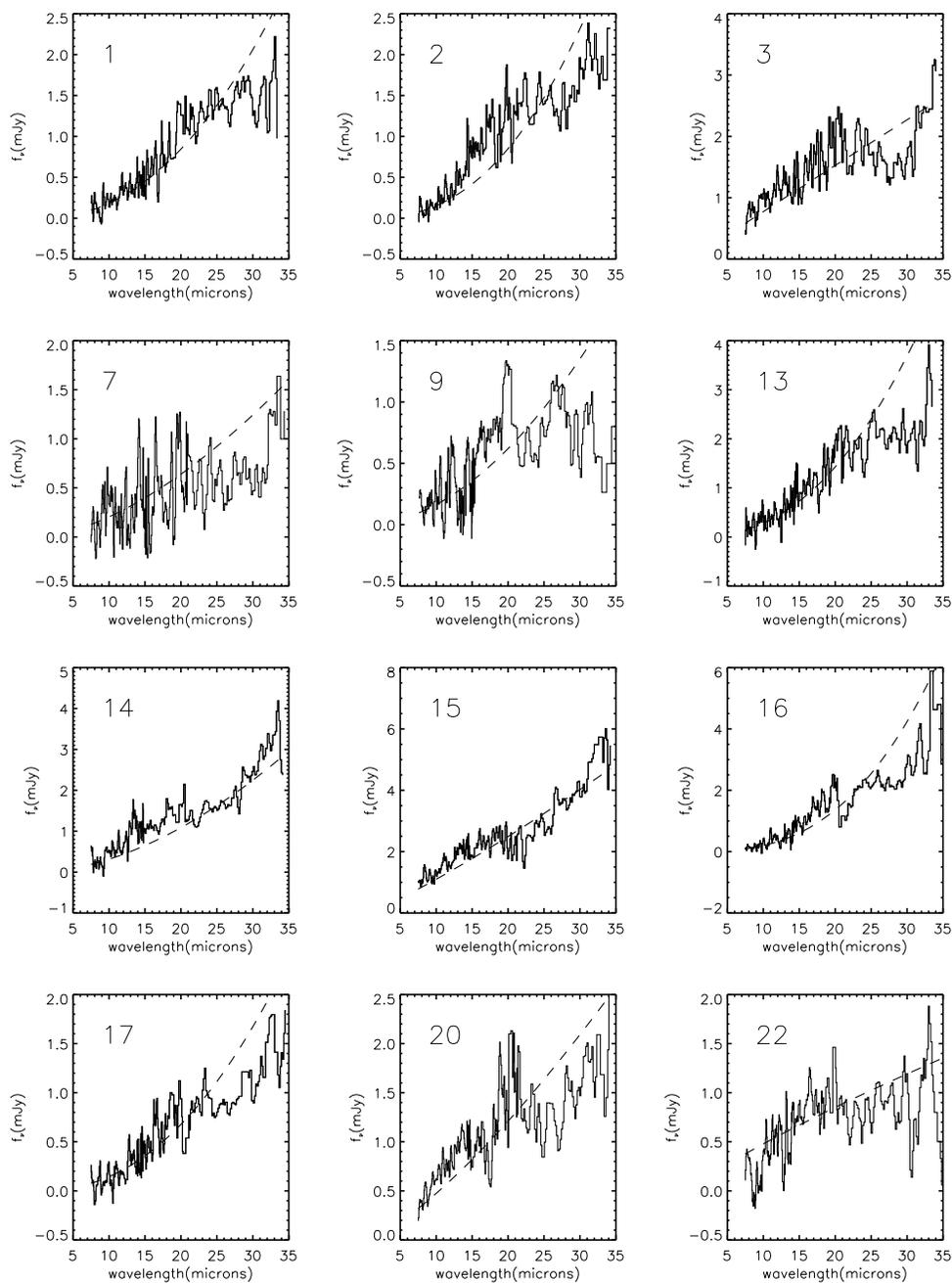}
\caption{Observed spectra of sources in Table 1 which have possible spectral features discussed in the text (histogram).  Spectra are smoothed to approximate resolution of individual IRS orders; dashed curve: power law that would connect MIPS f$_{\nu}$(24 \ums) with the IRAC f$_{\nu}$(8 \ums).}

\end{figure}

\newpage
\clearpage

\begin{figure}
  \figurenum{2}
\includegraphics[scale=0.7]{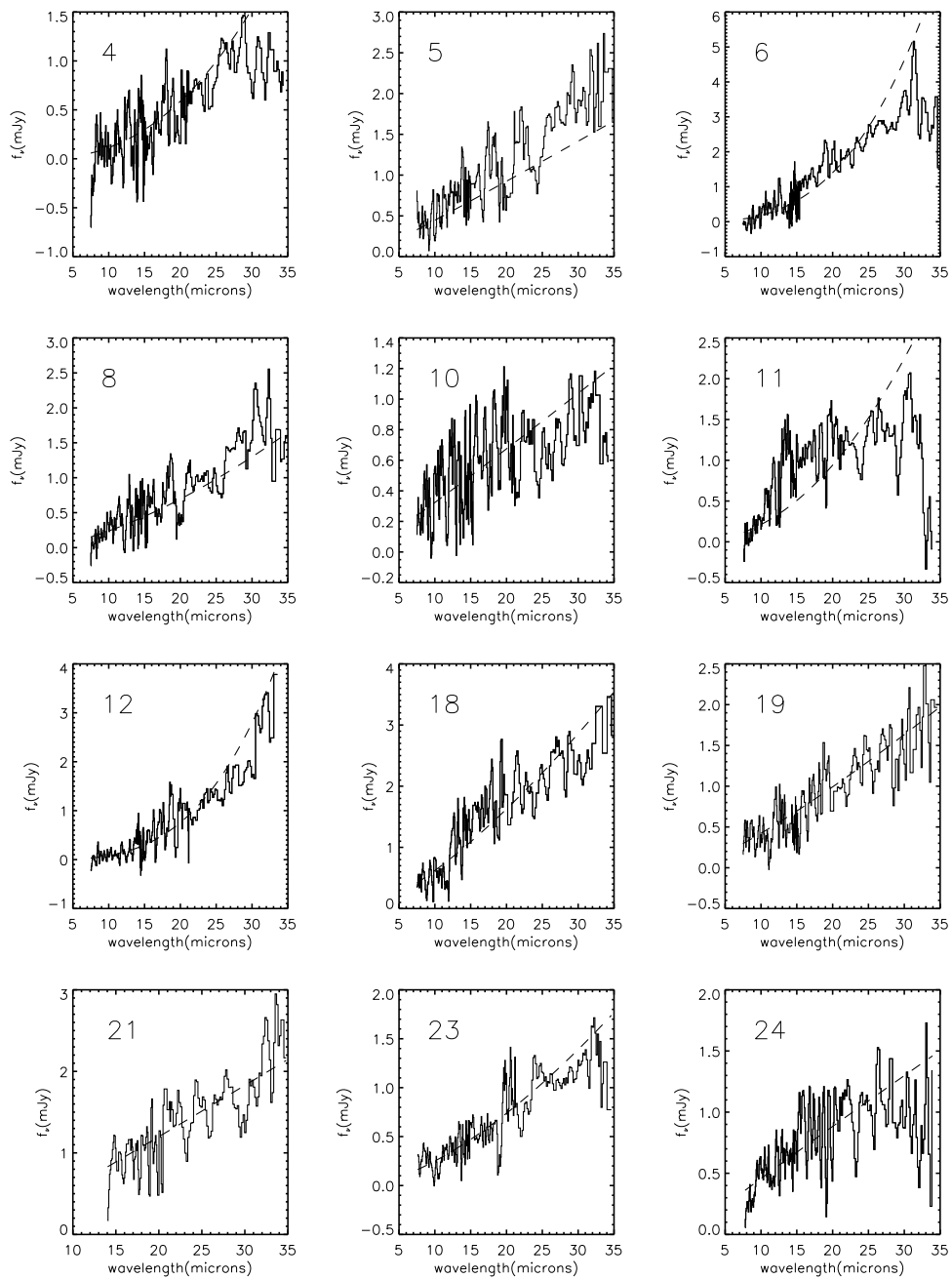}
\caption{Observed IRS spectra of featureless sources in Table 1 (histogram).  Spectra are smoothed to approximate resolution of individual IRS orders; dashed curve: power law that would connect MIPS f$_{\nu}$(24 \ums) with the IRAC f$_{\nu}$(8 \ums).}
\end{figure}

\newpage
\clearpage

\begin{figure}
  \figurenum{3}
\includegraphics[scale=0.9]{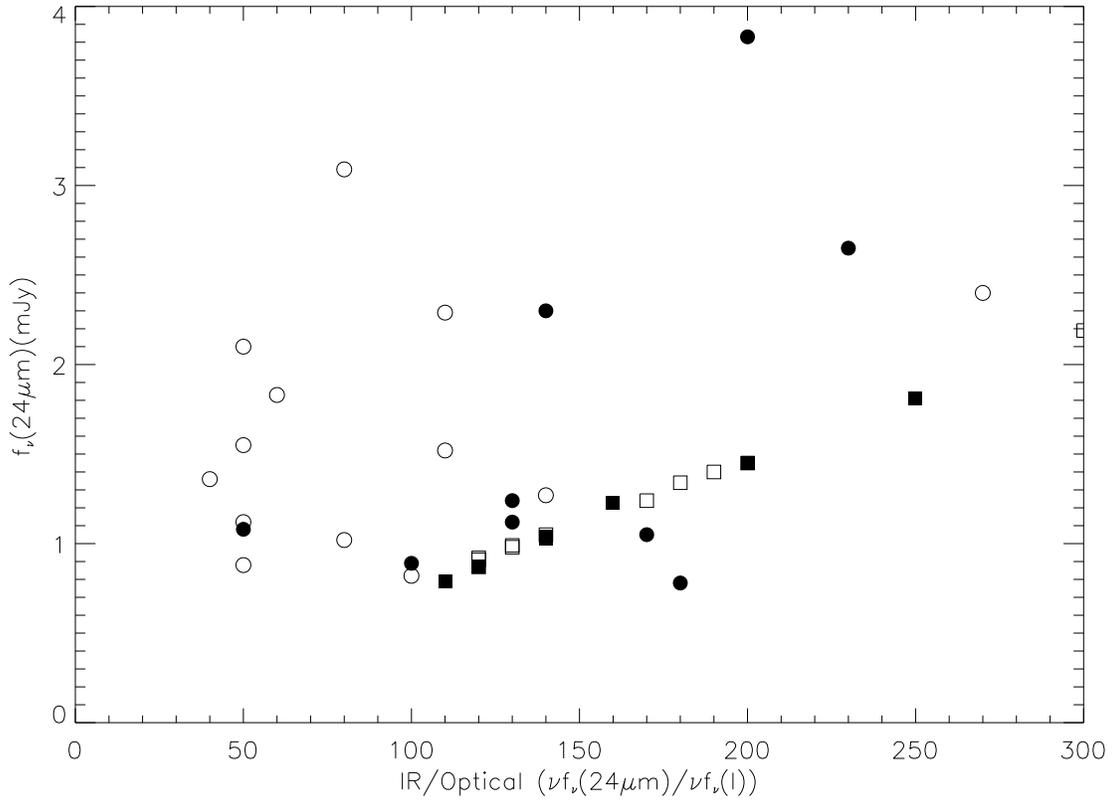}
\caption{Values of f$_{\nu}$(24\,\um) and IR/opt $= \nu$f$_{\nu}$(24\,\um)$/ \nu$f$_{\nu}$($I$) for sources in Table 1; measured IR/opt shown as open circles and lower limits on IR/opt shown as open squares. For comparison, 24\,\um flux densities and IR/opt are also shown for sources in H05, which have redshifts; measured IR/opt shown as filled circles and lower limits on IR/opt as filled squares.}
\end{figure}

\newpage
\clearpage

\begin{figure}
  \figurenum{4}
\includegraphics[scale=0.9]{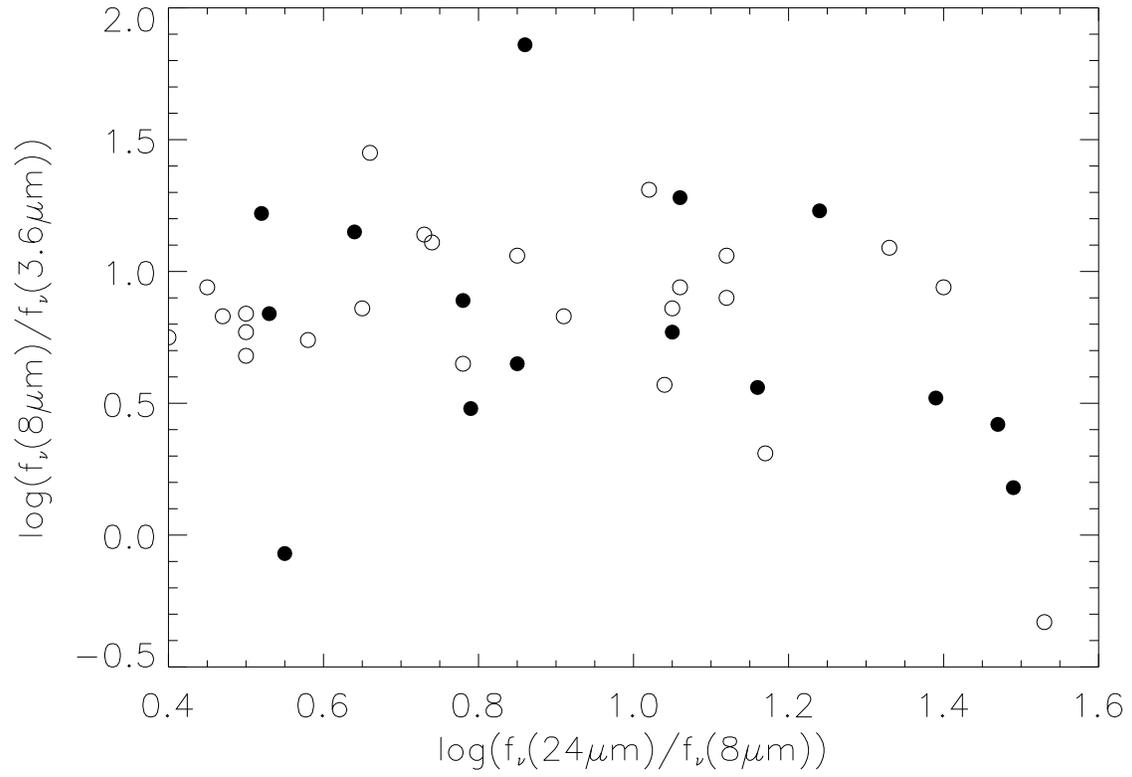}
\caption{Distributions of log[f$_{\nu}$(8\,\um)/f$_{\nu}$(3.6\,\um)] and log[f$_{\nu}$ (24\,\um)/f$_{\nu}$(8\,\um)] for sources in Table 1 (open circles); shown for comparison are sources with redshifts in H05 (filled circles).}
\end{figure}

\newpage
\clearpage

\begin{figure}
\figurenum{5}
\begin{center}
\includegraphics[scale=0.7, angle=90]{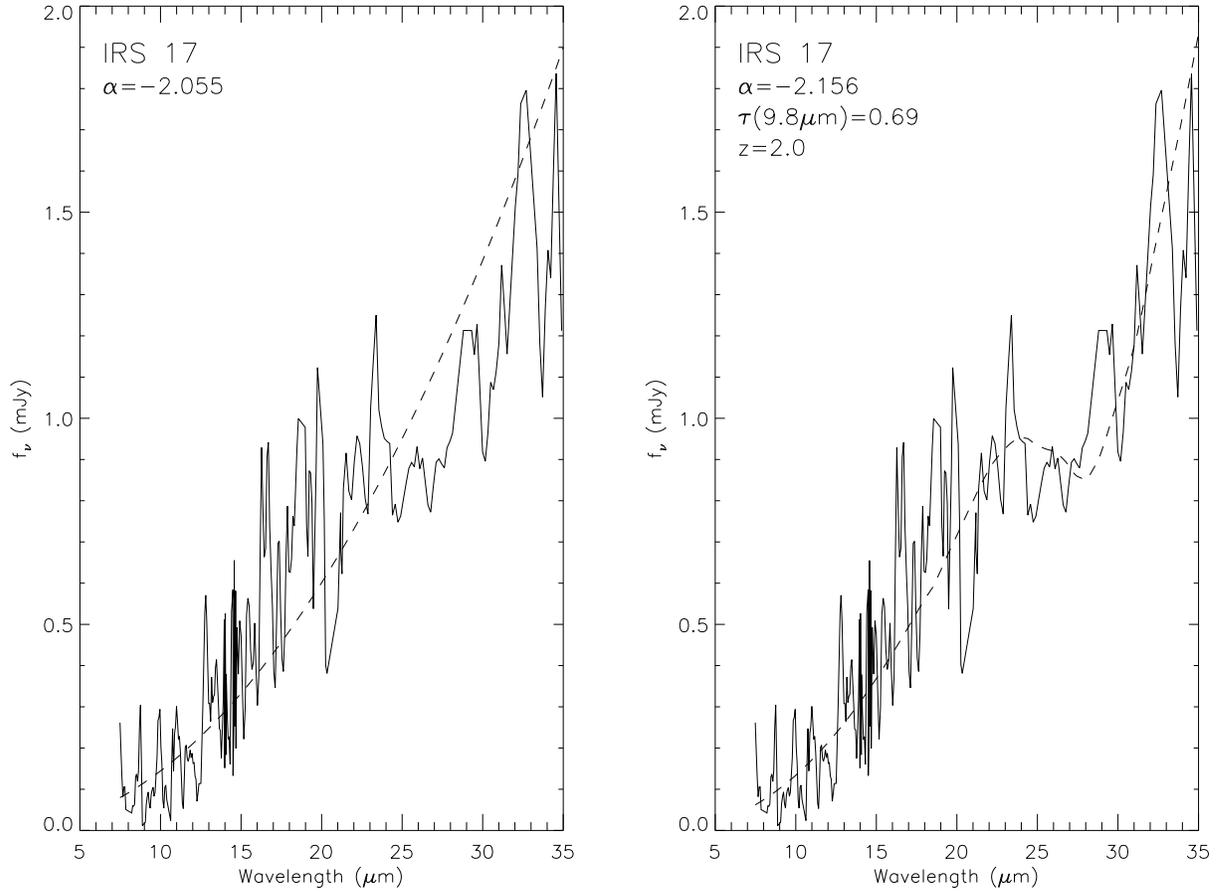}
\caption{Left panel: IRS spectrum of source 17 (solid curve) with power-law fit (dashed curve); Right panel: IRS spectrum of source 17(solid curve) with power-law fit plus a silicate absorption feature and redshift 2.0 (dashed curve).}
\end{center}
\end{figure}

\newpage
\clearpage

\begin{figure}
  \figurenum{6}
\includegraphics[scale=1.0]{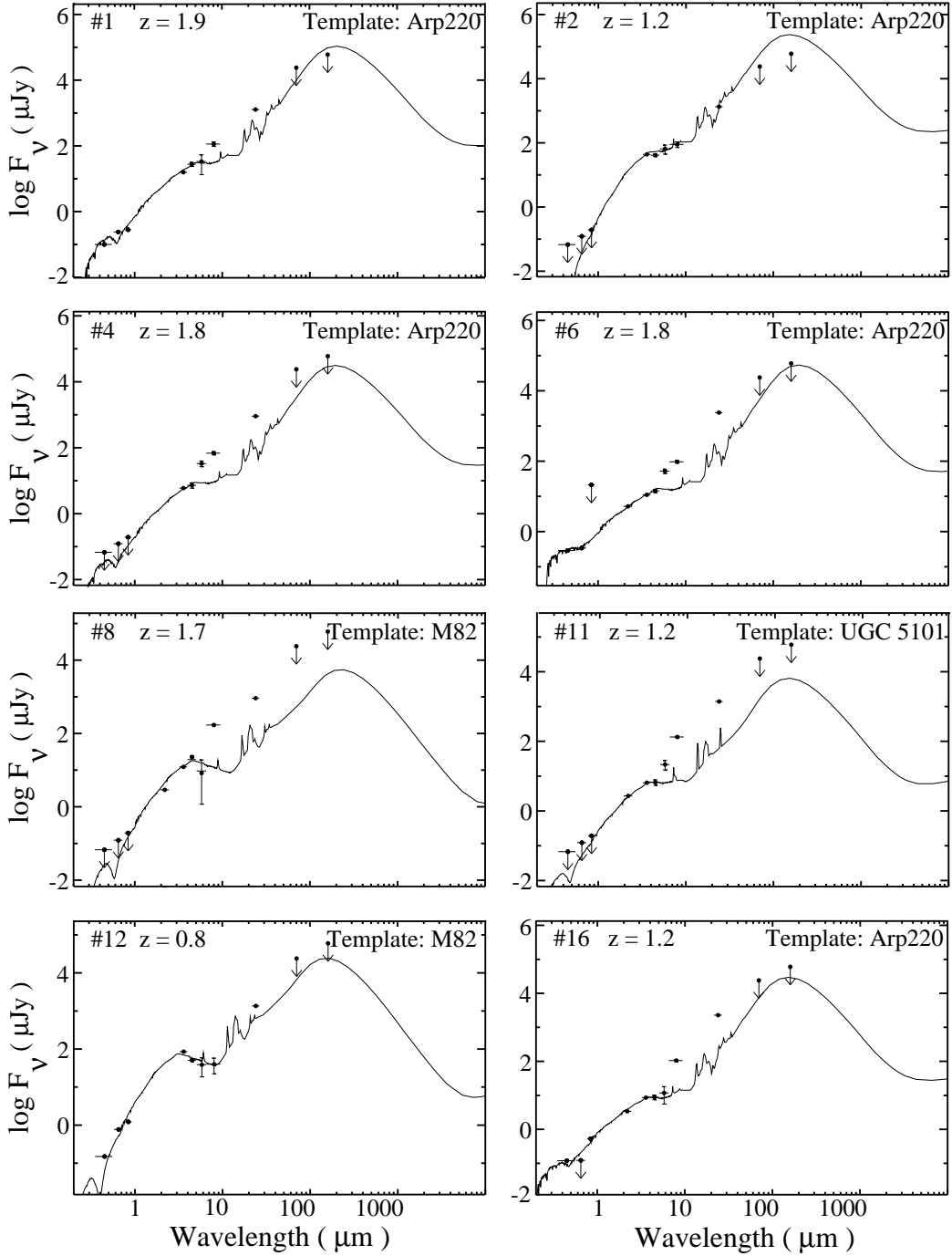}
\caption{Spectral energy distributions for sources in Table 2 which show IRAC fluxes consistent with a photospheric bump that allows a photometric redshift estimate; filled circles: observed photometric data points; solid curve: template fit which leads to value of photometric redshift shown in panel.}
\end{figure}

\end{document}